\documentclass[english]{article}
\usepackage{spconf}
\usepackage{array}
\usepackage[T1]{fontenc}
\usepackage[utf8]{inputenc}
\usepackage{geometry}
\usepackage{algorithmic}
\usepackage{varwidth}
\usepackage{listings}
\usepackage{amsmath}
\usepackage{graphicx}
\usepackage{float}
\usepackage{babel}
\newfloat{algorithm}{tbp}{loa}
\floatname{algorithm}{Algorithm}

\ninept
\twoauthors
{Steinar Midtskogen}{Cisco Systems Inc.\\Lysaker, Norway\\\href{mailto:stemidts@cisco.com}{stemidts@cisco.com}}
{Jean-Marc Valin}{Mozilla Corporation\\Mountain View, CA, USA\\\href{mailto:jmvalin@jmvalin.ca}{jmvalin@jmvalin.ca}}

\makeatletter

\newcolumntype{C}{>{\centering\arraybackslash}p{0.8em}}
\newcolumntype{X}{>{\centering\arraybackslash}p{10.5em}}

\providecommand{\tabularnewline}{\\}

\usepackage{color}
\usepackage{url}
\usepackage[pdfpagemode=None,pdfstartview=FitH,pdfview=FitH,colorlinks=true,pdftitle=The AV1 Constrained Directional Enhancement Filter,pdfauthor=Jean-Marc Valin]{hyperref}

\makeatother

\begin{document}

\title{The AV1 Constrained Directional Enhancement Filter (CDEF)}

\maketitle
\begin{abstract}
This paper presents the constrained directional enhancement filter
designed for the AV1 royalty-free video codec. The in-loop filter is based on
a non-linear low-pass filter and is designed for vectorization
efficiency. It takes into account the direction of edges and patterns
being filtered. The filter works by identifying the direction of each
block and then adaptively filtering with a high degree of control over
the filter strength along the direction and across it.  The proposed
enhancement filter is shown to improve the quality of the
Alliance for Open Media (AOM) AV1 and Thor video codecs in particular in low
complexity configurations.
\end{abstract}

\begin{keywords}enhancement filter, directional filter, AV1\end{keywords}

\section{Introduction}
\label{sec:intro}

The main goal of the in-loop constrained directional enhancement filter (CDEF)
is to filter out coding artifacts while
retaining the details of the image. In HEVC~\cite{sullivan2012overview}, the
Sample Adaptive Offset~(SAO)~\cite{HEVC-SAO} algorithm achieves a similar goal by defining
signal offsets for different classes of pixels. Unlike SAO, the
approach we take in AV1 is that of a non-linear spatial filter.  From
the very beginning, the design of the filter was constrained to be
easily vectorizable (i.e. implementable with SIMD operations), which
was not the case for other non-linear filters like the median
filter and the bilateral filter~\cite{Bilateral}.

The CDEF design originates from the following observations.
The amount of ringing artifacts in a coded image tends to be roughly proportional
to the quantization step size. The amount of detail is a property
of the input image, but the smallest detail actually retained in the
quantized image tends to also be proportional to the quantization
step size. For a given quantization step size, the amplitude of the
ringing is generally less than the amplitude of the details.

CDEF works by identifying the
direction~\cite{DaedeDCC} of each block (Sec.~\ref{sec:direction-search}) and then adaptively filtering along the
identified direction (Sec.~\ref{sec:nonlinear-lowpass}) and to a lesser degree along directions rotated
45~degrees from the identified direction.  The filter strengths are signaled explicitly, which
allows a high degree of control over the blurring (Sec.~\ref{sec:signaling}).
Sec.~\ref{sec:encoder-search} demonstrates an efficient encoder search for the filter strengths, with
results presented in Sec.~\ref{sec:results}.

\section{Direction Search}
\label{sec:direction-search}

The direction search operates on the
reconstructed pixels, just after the deblocking filter. Since
those pixels are available to the decoder, the directions require no signaling.
The search operates on $8\times8$ blocks, which are
small enough to adequately handle non-straight edges,
while being large enough to reliably estimate directions when applied
to a quantized image. Having a constant direction over an $8\times8$
region also makes vectorization of the filter easier.

\begin{figure}
\centering{\includegraphics[width=0.24\columnwidth]{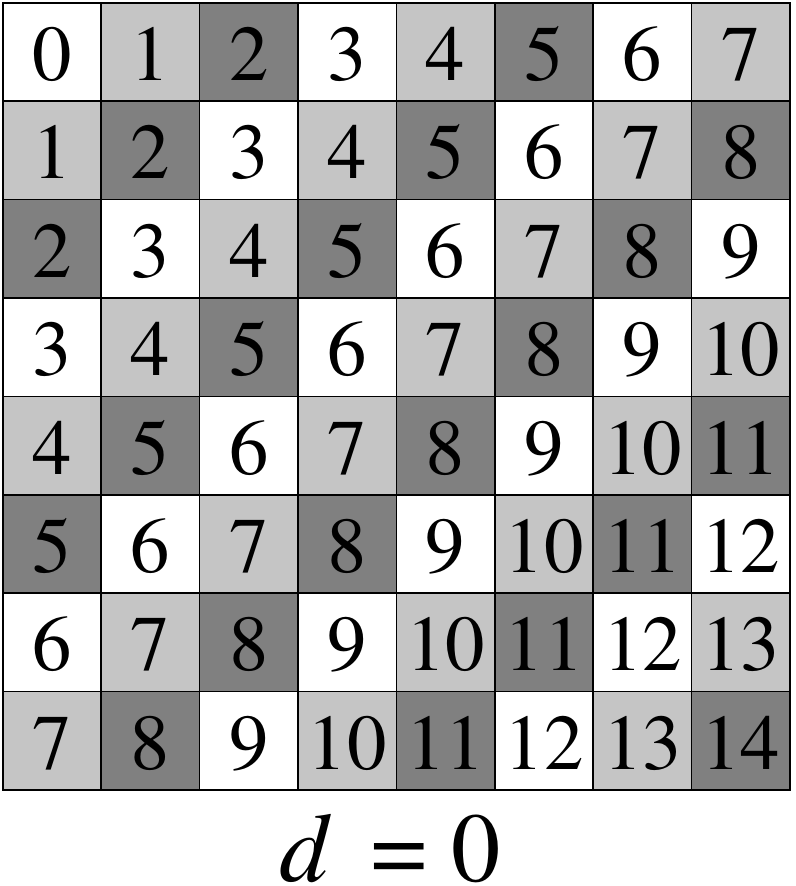}\hspace{.35em}\includegraphics[width=0.24\columnwidth]{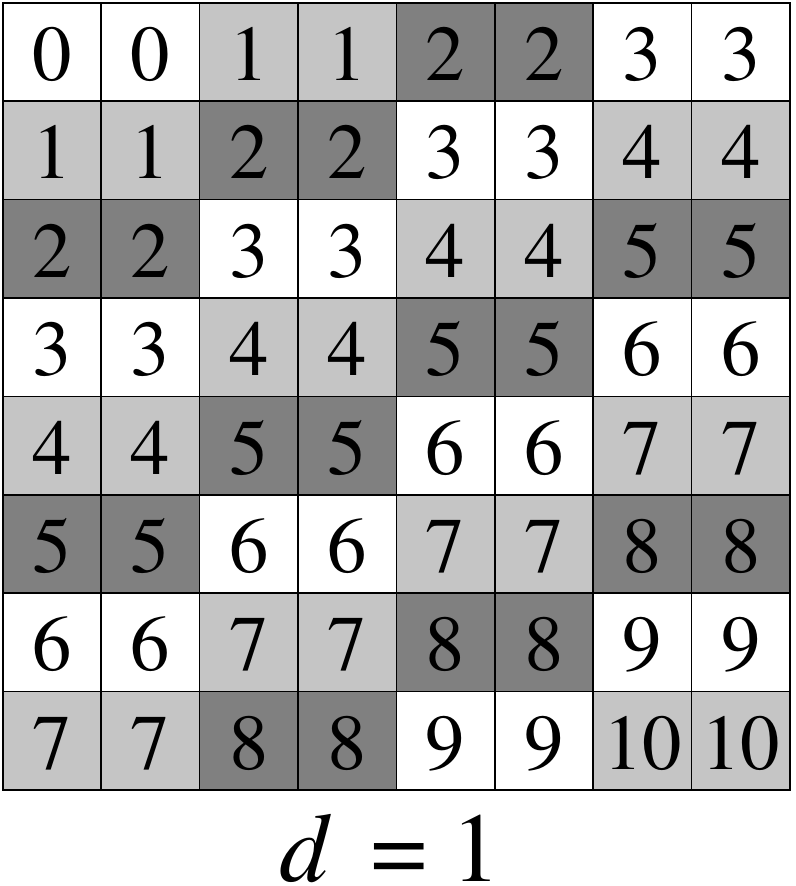}\hspace{.35em}\includegraphics[width=0.24\columnwidth]{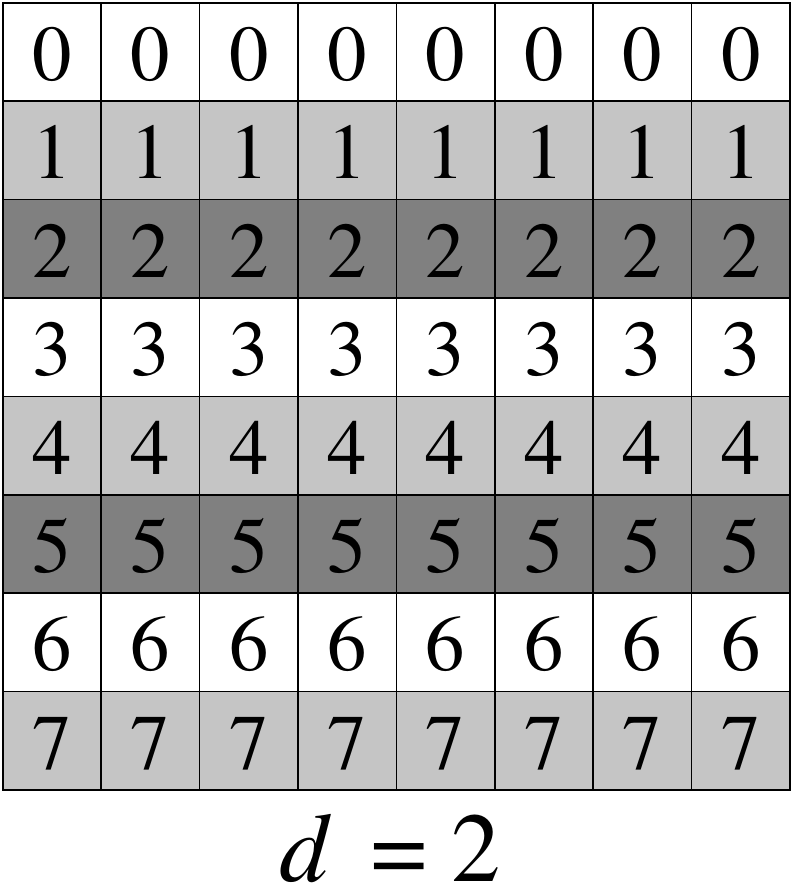}\hspace{.35em}\includegraphics[width=0.24\columnwidth]{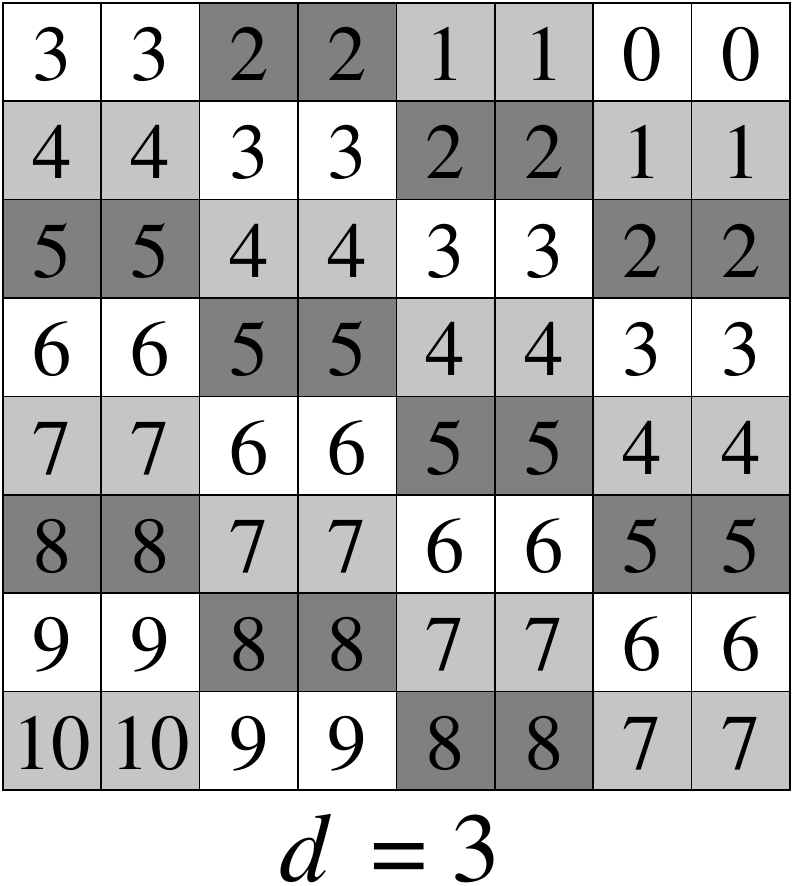}}

\vspace{.5em}

\centering{\includegraphics[width=0.24\columnwidth]{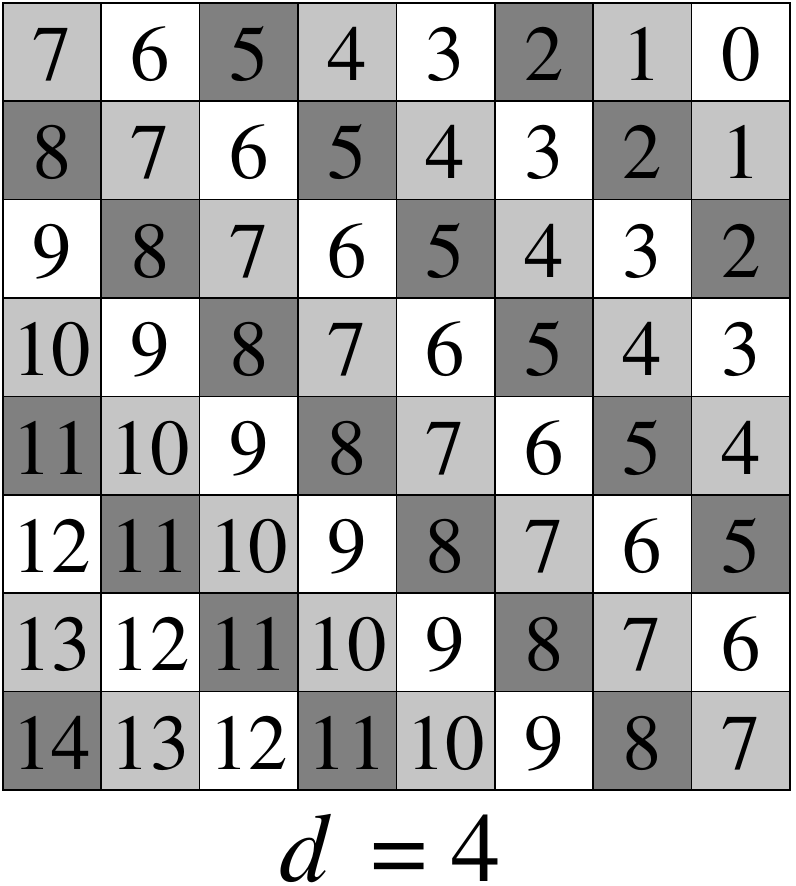}\hspace{.35em}\includegraphics[width=0.24\columnwidth]{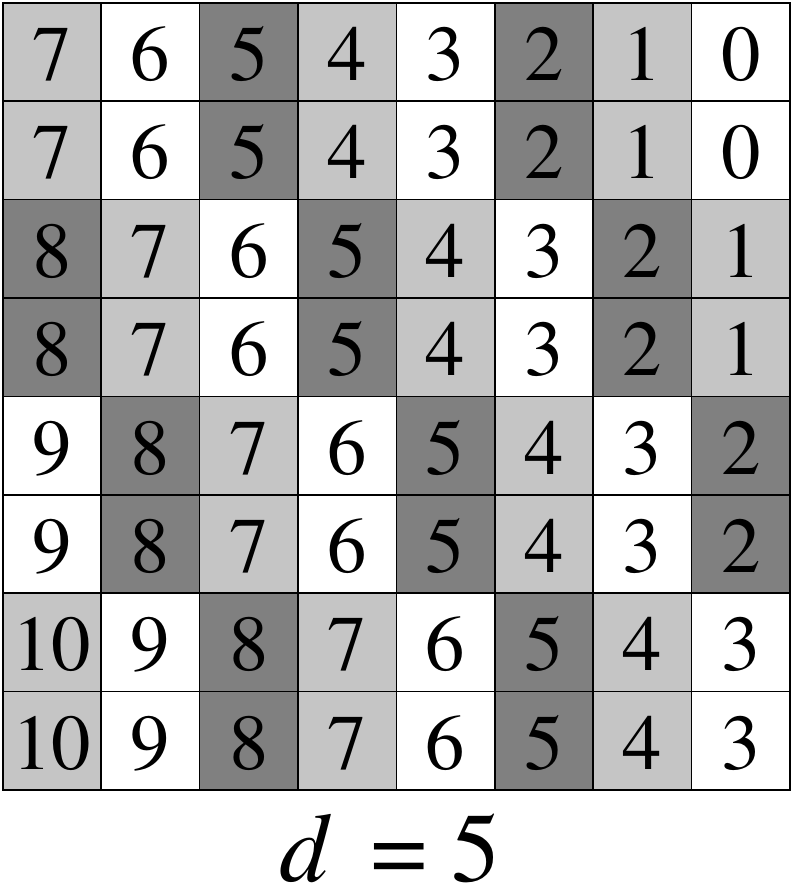}\hspace{.35em}\includegraphics[width=0.24\columnwidth]{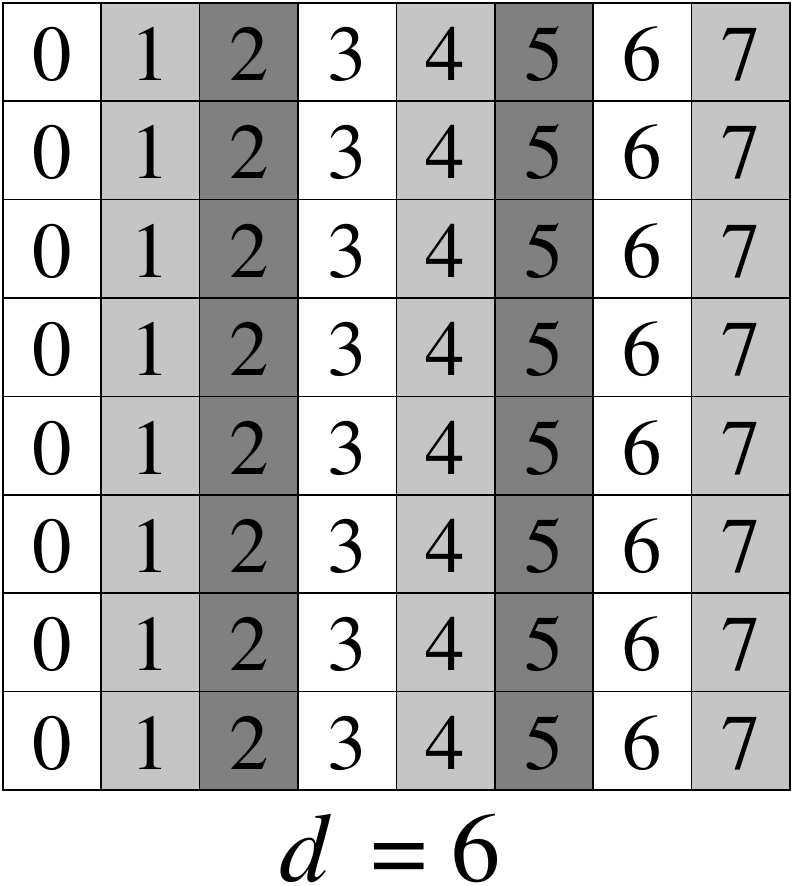}\hspace{.35em}\includegraphics[width=0.24\columnwidth]{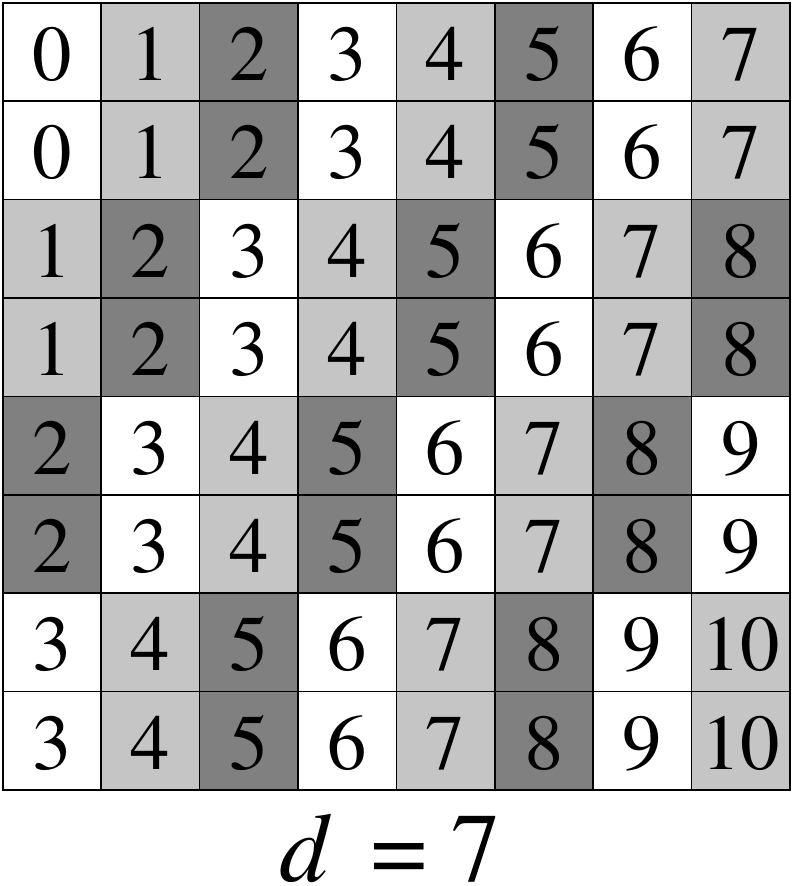}}\caption{Line number $k$ for pixels following direction $d=0:7$ in an $8\times8$
block.\label{fig:Lines-for-direction}}
\end{figure}

For each block we determine the direction that best matches
the pattern in the block by minimizing the sum of squared
differences (SSD) between the quantized block and the closest perfectly directional
block. A perfectly directional block is a block where all of the pixels along
a line in one direction have the same value. For each direction,
we assign a line number $k$ to each pixel, as shown in Fig.~\ref{fig:Lines-for-direction}. 

For each direction $d$, the pixel average for line $k$ is
\begin{equation}
\mu_{d,k}=\frac{1}{N_{d,k}}\sum_{p\in P_{d,k}}x_{p}\ ,\label{eq:pixel-average}
\end{equation}
where $x_{p}$ is the value of pixel $p$, $P_{d,k}$ is the set of
pixels in line $k$ following direction $d$ and $N_{d,k}$ is the
cardinality of $P_{d,k}$ (for example, in Fig.~\ref{fig:Lines-for-direction},
$N_{1,0}=2$ and $N_{1,4}=8$). The SSD is then
\begin{equation}
E_{d}^{2}=\sum_{k}\left[\sum_{p\in P_{d,k}}\left(x_{p}-\mu_{d,k}\right)^{2}\right]\ .\label{eq:direction-variance0}
\end{equation}

Substituting (\ref{eq:pixel-average}) into (\ref{eq:direction-variance0}) and
simplifying results in
\begin{equation}
E_{d}^{2}= \sum_{p}x_{p}^{2}-\sum_{k}\frac{1}{N_{d,k}}\left(\sum_{p\in P_{d,k}}x_{p}\right)^{2}\ .\label{eq:direction-variance1}
\end{equation}
 Note that the simplifications leading to (\ref{eq:direction-variance1})
are the same as to those allowing a variance to be computed as $\sigma_{x}^{2}=\frac{\sum x^{2}}{N}-\frac{\left(\sum x\right)^{2}}{N^2}$.
Considering that the first term of (\ref{eq:direction-variance1})
is constant with respect to $d$, we find the optimal direction
$d_{opt}$ by maximizing the second term:
\begin{align}
d_{opt}&=\max_{d}s_{d}\label{eq:direction-variance2}\\
s_{d}&=\sum_{k}\frac{1}{N_{d,k}}\left(\sum_{p\in P_{d,k}}x_{p}\right)^{2}\ .\label{eq:direction-variance3}
\end{align}

We can avoid the division in (\ref{eq:direction-variance3}) by multiplying
$s_{d}$ by 840, the least common multiple of the possible $N_{d,k}$
values ($1\le N_{d,k}\le8$). When using 8\nobreakdash-bit pixel
data, and centering the values such that $-128\le x_{p}\le127$, then $840s_{d}$ and
all calculations needed for $s_{d}$ fit in a 32\nobreakdash-bit
signed integer. For higher bit depths, we downscale the pixels to $8$ bits
during the direction search.

\begin{figure}
\centering{\includegraphics[width=1.0\columnwidth]{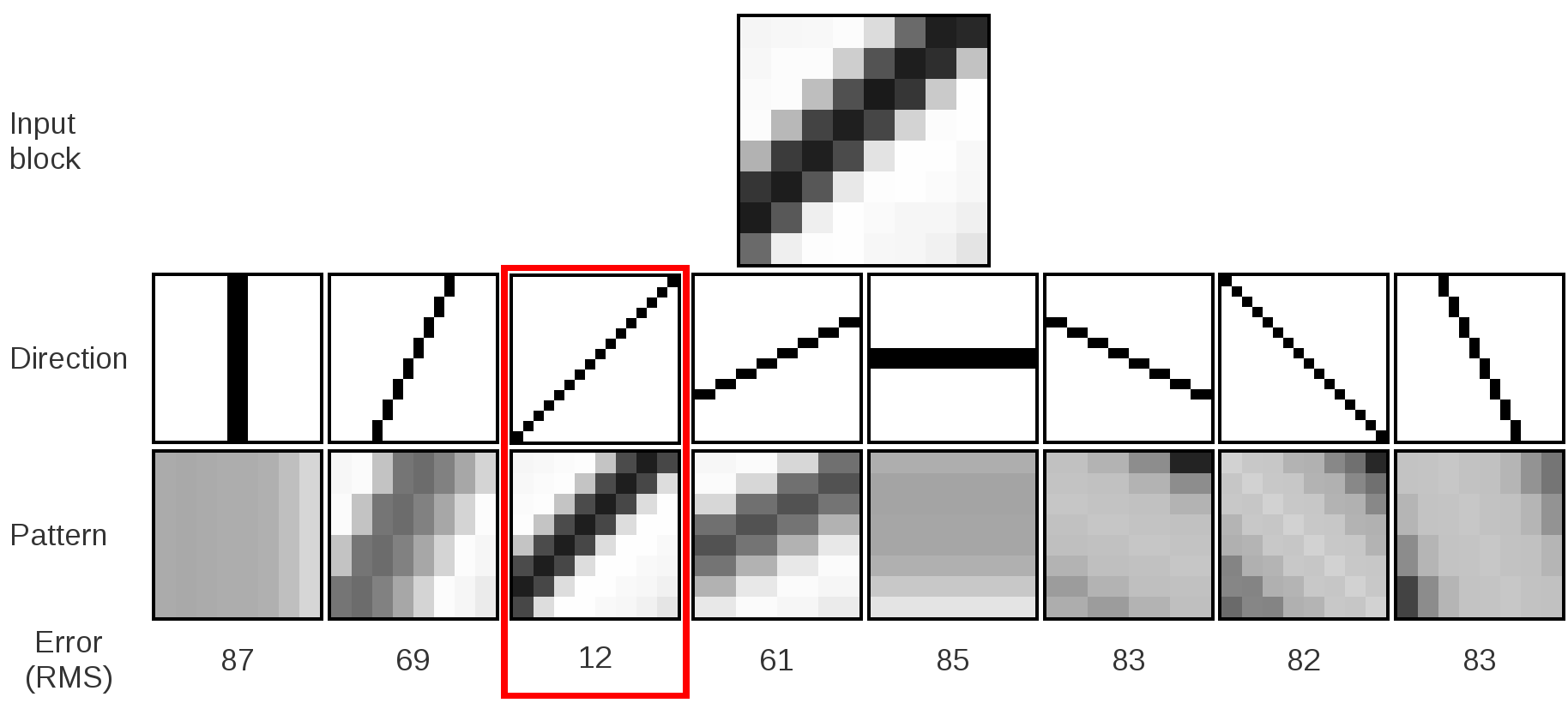}}\caption{Example of direction search for an $8\times8$ block. The patterns
shown are based on the $\mu_{d,k}$ values. In this case, the 45-degree
direction is selected because it minimizes $E_{d}^{2}$. The error values $E_{d}$
shown are never computed in practice (only $s_{d}$ is). \label{fig:Example-of-direction}}

\end{figure}

Fig. \ref{fig:Example-of-direction} shows an example of a direction
search for an $8\times8$ block containing a line. 
The step-by-step process is described in algorithm~\ref{cap:direction-search}.
To save on decoder complexity, we assume that luma and chroma directions are correlated,
and only search the luma component. The same direction is used for the chroma components.

In total, the search for all 8 directions requires the following
arithmetic operations:
\begin{enumerate}
\item The pixel accumulations in equation (\ref{eq:direction-variance3})
can be implemented with 294 additions (reusing partial sums of adjacent pixels).
\item The accumulations result in 90 line sums. Each is squared, requiring 90 multiplies.
\item The $s_d$ values can be computed from the squared line sums with 34 multiplies and 82 additions.
\item Finding the largest $s_d$ value requires 7 comparisons.
\end{enumerate}

The total is 376~additions, 124~multiplies and 7 comparisons. That is about two thirds of the number of operations required
for the 8x8 IDCT in HEVC~\cite{Budagavi2014}. The code can be efficiently vectorized,
with a small penalty due to the diagonal alignment, resulting in a complexity similar to
that of an 8x8 IDCT.

\begin{algorithm}[t]
\noindent\fbox{%
\begin{varwidth}{\dimexpr\linewidth-2\fboxsep-2\fboxrule\relax}
\begin{algorithmic}
\STATE {Initialize all variables to zero}
\FOR{$d=0$ to $7$}
  \FOR{$i=0$ to $7$}
    \FOR{$j=0$ to $7$}
      \STATE $L \leftarrow \mathrm{line\_table}[d][i][j]$
      \STATE $\mathrm{partial}[d][L] \leftarrow \mathrm{partial}[d][L] + \left(\mathrm{pixel}[i][j] - 128\right)$
      \STATE $\mathrm{count}[d][L] \leftarrow \mathrm{count}[d][L] + 1$
    \ENDFOR
  \ENDFOR
  \FOR{$L=0$ to $14$}
    \IF{$\mathrm{count}[d][L] > 0$}
      \STATE $s[d] \leftarrow s[d] + \mathrm{partial}[d][L]^2\cdot 840/\mathrm{count}[d][L]$
    \ENDIF
  \ENDFOR
\ENDFOR
\FOR{$d=0$ to $7$}
  \IF{$s[d] > s[\mathrm{best\_}d]$}
    \STATE $\mathrm{best\_}d \leftarrow d$ 
  \ENDIF
\ENDFOR
\STATE $\mathrm{direction} \leftarrow \mathrm{best\_}d$
\STATE $\mathrm{directional\_contrast} \leftarrow s[\mathrm{best\_}d] - s[(\mathrm{best\_}d+4) \mod 8]$
\end{algorithmic}
\end{varwidth}%
}
\caption{Direction search. The line\_table[$d$][$i$][$j$] values are the line numbers shown
in Fig.~\ref{fig:Lines-for-direction}.
The $840/\mathrm{count}[d][L]$ terms can be pre-computed.
More functionally equivalent algebraic simplifications
are possible, but they are not shown here for clarity.\label{cap:direction-search}}
\end{algorithm}

\section{Non-linear Low-pass Filter}

\begin{figure}
\centering{\includegraphics[width=0.9\columnwidth]{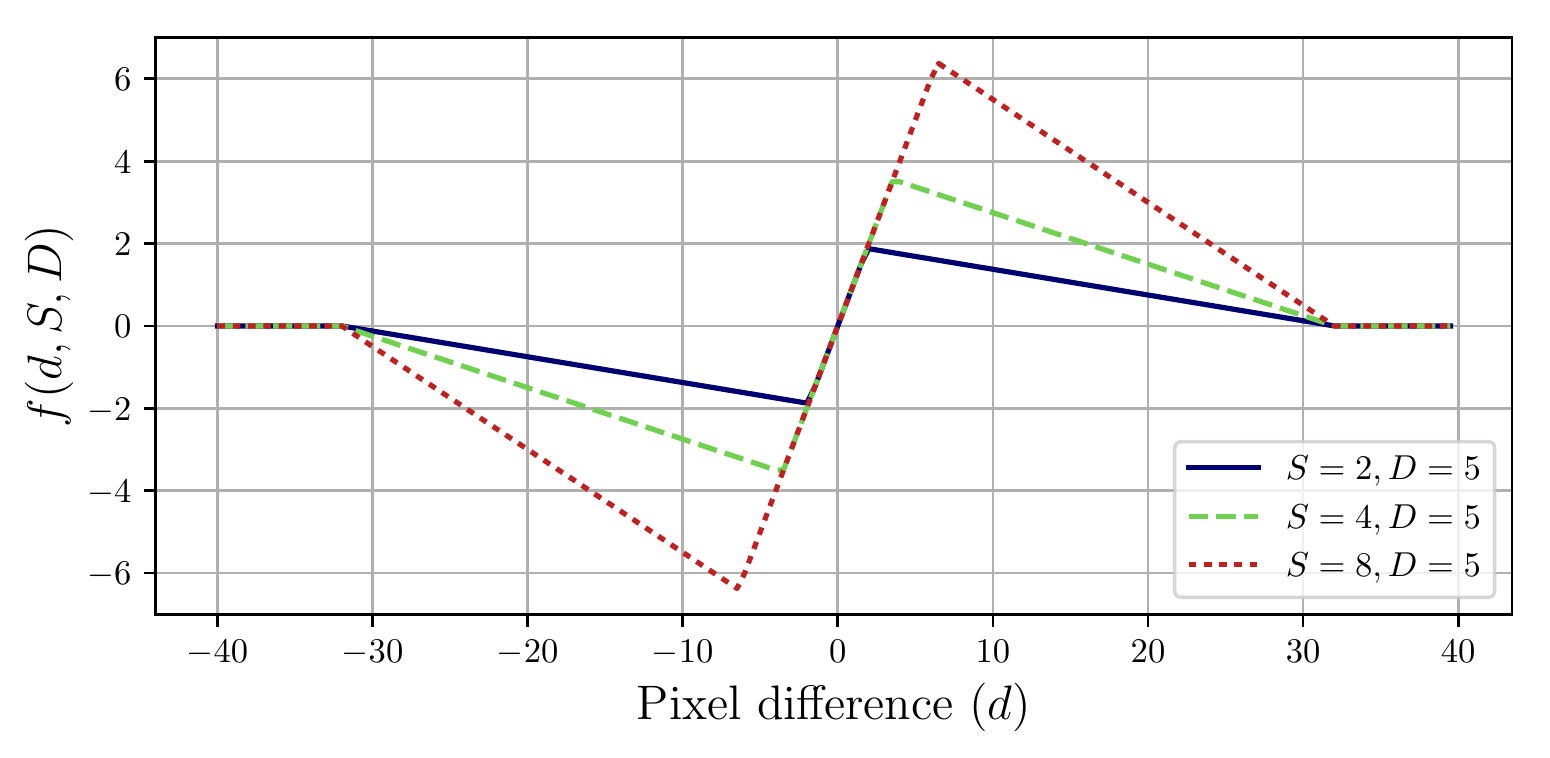}}
\centering{\includegraphics[width=0.9\columnwidth]{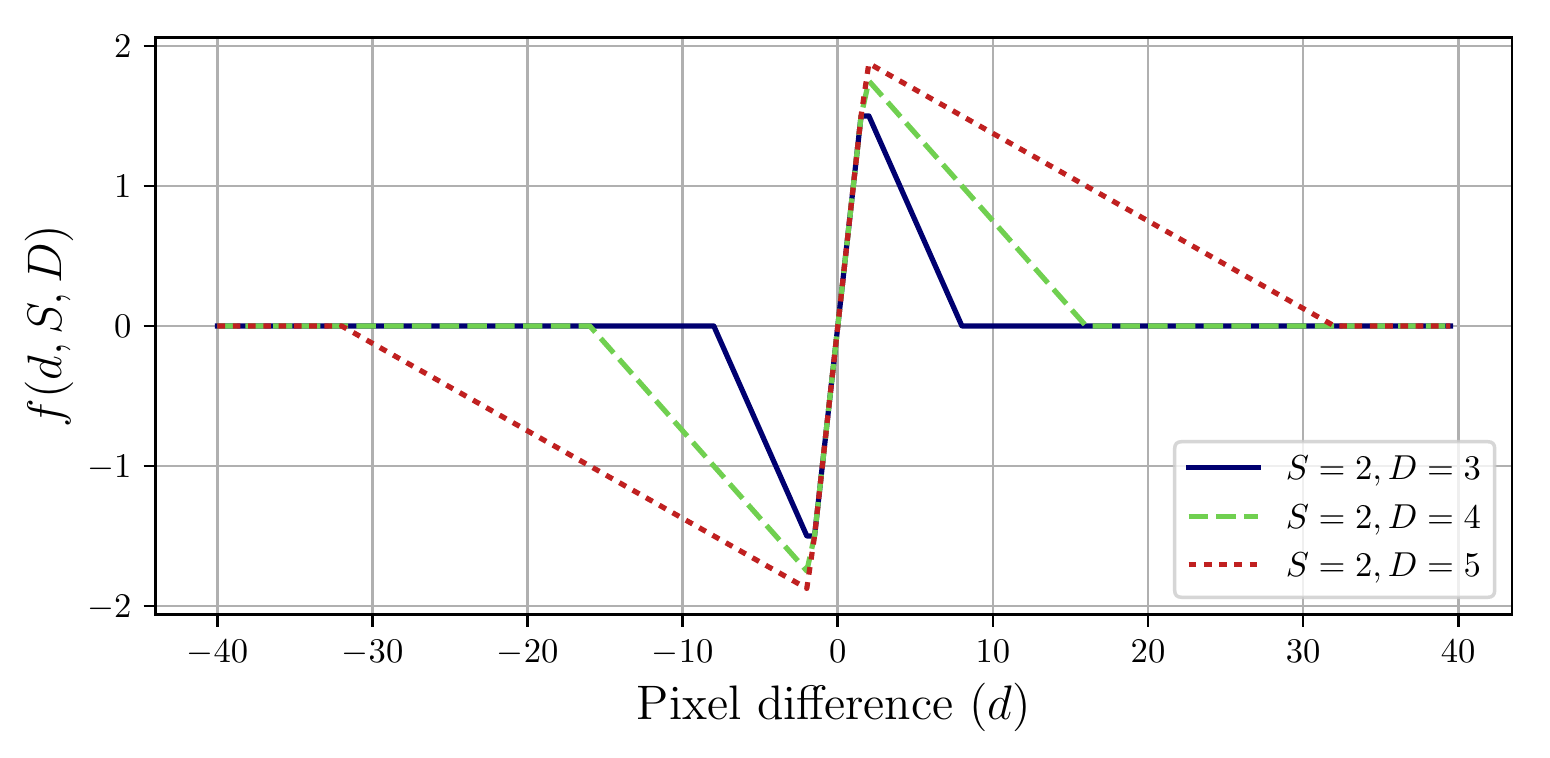}}
\vspace{-1.em}
\caption{Effect of strength (top) and damping (bottom) on $f(d, S, D)$.\label{fig:constraint-function-strength-damping}}
\end{figure}

\label{sec:nonlinear-lowpass}
CDEF uses a non-linear low-pass filter designed to remove coding artifacts without
blurring sharp edges. It achieves this by selecting filter tap locations based on the
identified direction, but also by preventing excessive blurring when the filter
is applied across an edge. The latter is achieved through the use of a non-linear low-pass
filter that deemphasizes pixels that differ too much from the pixel being filtered~\cite{Midtskogen2017}.
In one dimension, the non-linear filter is expressed as
\begin{equation}
y\left(i\right)=x\left(i\right) +\!\!\sum_{m}w_k f\left(x\left(i+m) - x(i\right), S, D\right)\ ,\label{eq:linear-filter}
\end{equation}
where $w_k$ are the filter weights and $f(d, S, D)$ is a \textit{constraint function} operating
on the difference between the filtered pixel and each of the
neighboring pixels. For small differences, $f\left(d, S, D\right)=d$, making the filter in
(\ref{eq:linear-filter}) behave like a linear filter. When the difference is large,
$f\left(d, S, D\right)=0$, which effectively ignores the filter tap.
The filter is parametrized by a \textit{strength} $S$ and a \textit{damping} $D$:
\begin{equation}
f\left(d, S, D\right)=\left\{\begin{array}{ll}
\!\!\min\left(d, \max\left(0, S-\left\lfloor{\frac{d}{2^{D-\lfloor{\log_{2}S}\rfloor}}}\right\rfloor\right)\right) , d\ge0\\
\!\!\max\left(d, \min\left(0, \left\lceil{\frac{-d}{2^{D-\lfloor{\log_{2}S}\rfloor}}}\right\rceil-S\right)\right) , d<0
\end{array}\right. \label{eq:constraint-function}
\end{equation}
with $D \ge \lfloor\log_2{S}\rfloor$.
The strength $S$ controls the maximum difference allowed and the damping $D$ controls the point where
$f\left(d, S, D\right)=0$.  Fig.~\ref{fig:constraint-function-strength-damping} illustrates
the effect of the strength and damping on $f(\cdot)$. The function is anti-symmetric around $d=0$.

\subsection{Directional filter}

\begin{figure}
\centering{\includegraphics[width=0.24\columnwidth]{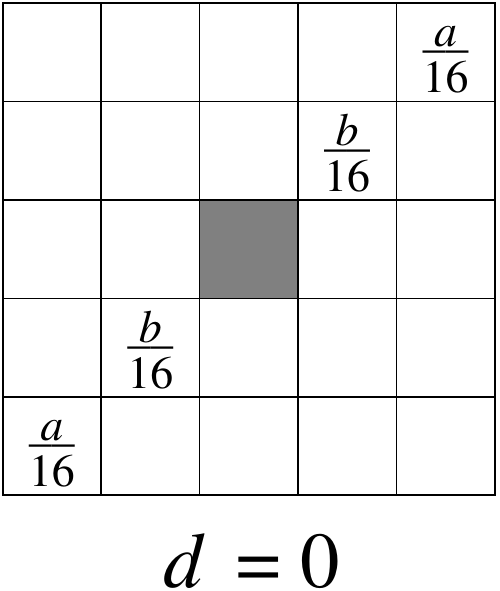}\hspace{.35em}\includegraphics[width=0.24\columnwidth]{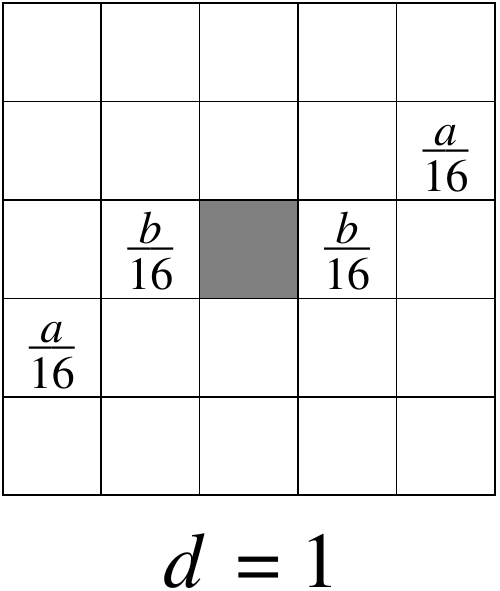}\hspace{.35em}\includegraphics[width=0.24\columnwidth]{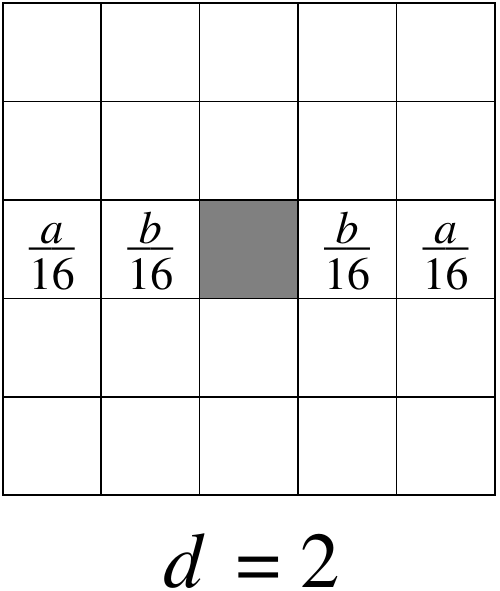}\hspace{.35em}\includegraphics[width=0.24\columnwidth]{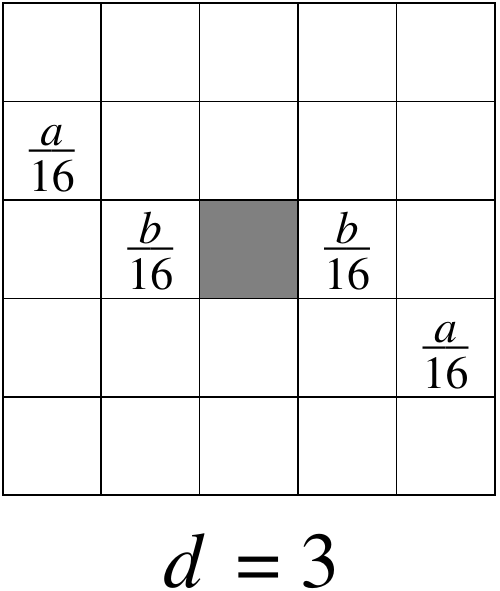}}

\vspace{.5em}

\centering{\includegraphics[width=0.24\columnwidth]{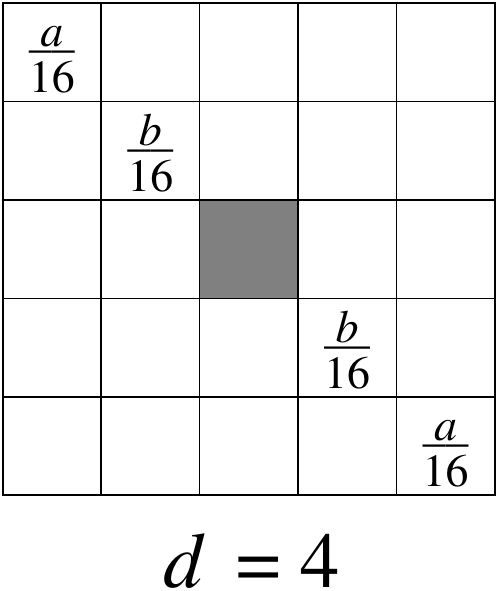}\hspace{.35em}\includegraphics[width=0.24\columnwidth]{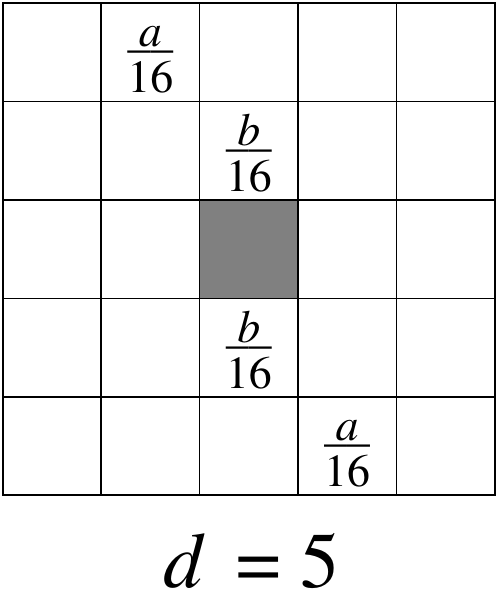}\hspace{.35em}\includegraphics[width=0.24\columnwidth]{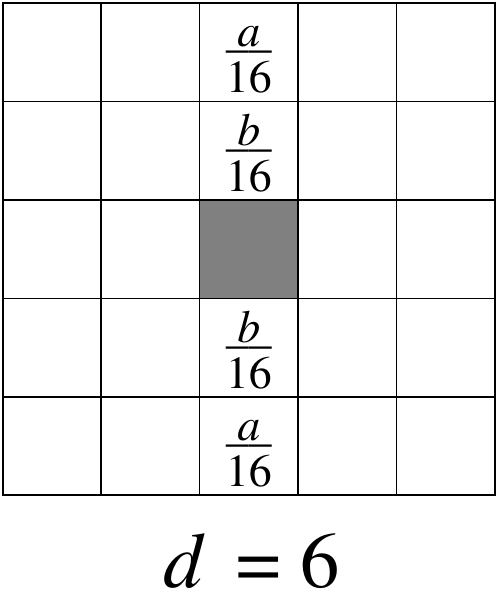}\hspace{.35em}\includegraphics[width=0.24\columnwidth]{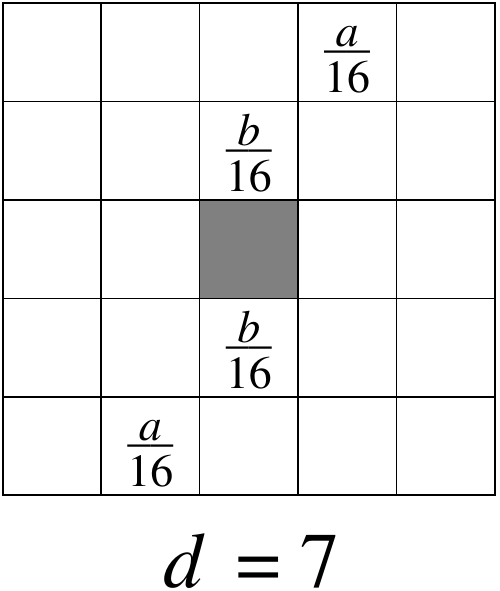}}
\vspace{-1.5em}
\caption{Primary filter taps following direction $d=0:7$. For even strengths $a=2$ and $b=4$, whereas for odd strengths $a=3$ and $b=3$.
The filtered pixel in shown in gray.
\label{fig:Primary-filter}}
\end{figure}

\begin{figure}
\centering{\includegraphics[width=0.24\columnwidth]{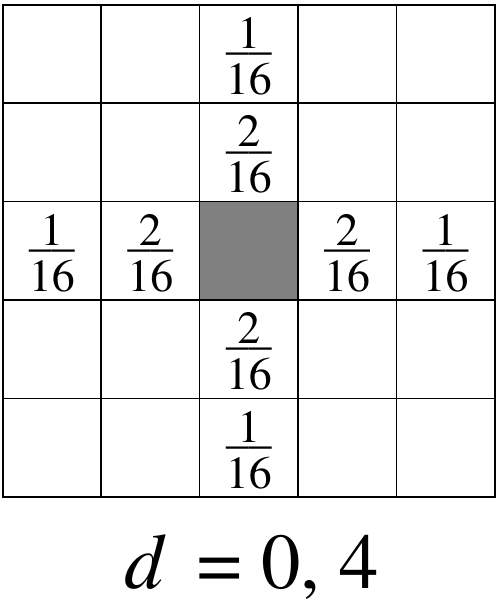}\hspace{.35em}\includegraphics[width=0.24\columnwidth]{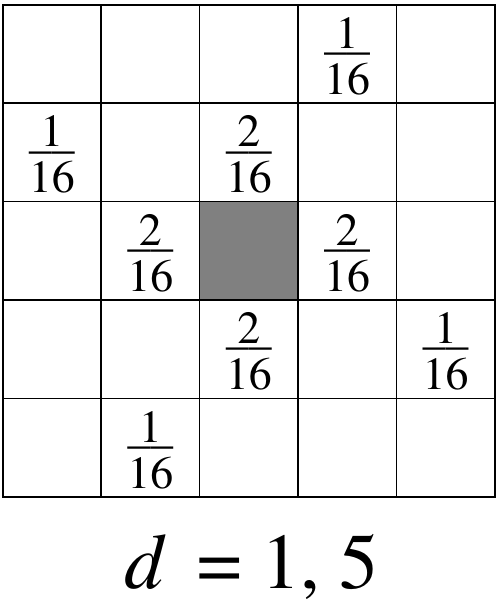}\hspace{.35em}\includegraphics[width=0.24\columnwidth]{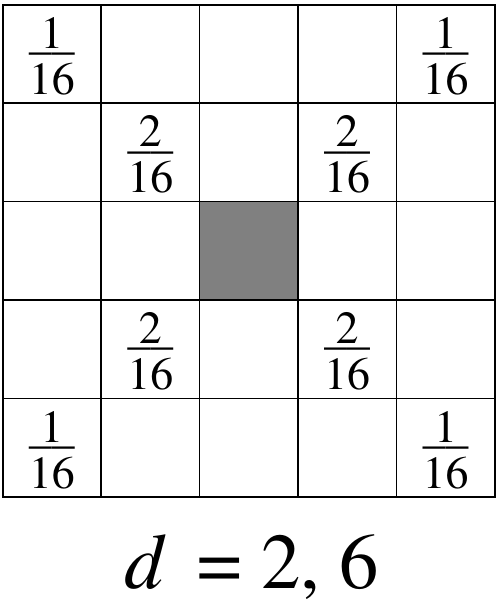}\hspace{.35em}\includegraphics[width=0.24\columnwidth]{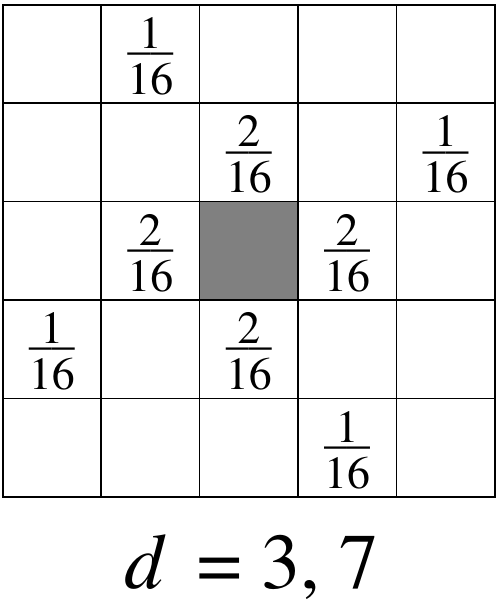}}
\vspace{-1.5em}
\caption{Secondary filter taps following direction $d=0:7$. The filtered pixel is shown in gray.\label{fig:Secondary-filter}}
\end{figure}

The main reason for identifying the direction of the Section~\ref{sec:direction-search} is to
align the filter taps along that direction to reduce ringing while preserving the directional
edges or patterns. However, directional filtering alone sometimes cannot
sufficiently reduce ringing. We also want to use filter taps on pixels that do not
lie along the main direction.
To reduce the risk of blurring, these extra taps are treated more conservatively. For this
reason, CDEF defines \textit{primary taps} and \textit{secondary taps}. The primary taps follow
the direction $d$, and the weights are shown in Fig.~\ref{fig:Primary-filter}.  For the primary
taps, the weights alternate for every other strength, so that the weights for strengths 1, 3, 5,
etc. are different from the weights for strengths 2, 4, 6, etc. The secondary taps
form a cross, oriented $45^\circ$ off the direction $d$ and their weights are shown in
Fig.~\ref{fig:Secondary-filter}. The complete 2-D CDEF filter is expressed as
\begin{align}
y\left(i, j\right)=&x(i, j) + \mathrm{round}\bigg( \nonumber\\
 & \sum_{m,n}{w}^{(p)}_{d,m,n}f\left(x\left(m, n) - x(i,j\right), S^{(p)}, D\right) \nonumber\\
+& \sum_{m,n}{w}^{(s)}_{d,m,n}f\left(x\left(m, n) - x(i,j\right), S^{(s)}, D\right) \bigg)\ ,
\label{eq:linear-filter2}
\end{align}
where $S^{(p)}$ and $S^{(s)}$ and the strengths of the primary and secondary taps, respectively,
and $\mathrm{round}(\cdot)$ rounds ties away from zero.

Since the sum of all the primary and secondary weights exceed unity, it is possible (though rare) for
the output $y(i,j)$ to change by more than the maximum difference between the input and the neighboring
values. This is avoided by explicitly clamping the filter output based on the neighboring pixels with
non-zero weights:
\begin{align}
y_{clip}(i,j) &= \min\left(y_{max}, \max\left(y_{min}, y(i,j) \right) \right)\\
y_{min} &= \min_{m, n \in R}(x(i+m,j+n)) \\
y_{max} &= \max_{m, n \in R}(x(i+m,j+n)) \\
R &= (n,m) | {w}^{(p)}_{d,m,n}+{w}^{(s)}_{d,m,n} \ne 0
\label{eq:cap-function}
\end{align}

The direction, strength and damping parameters are constant over each $8\times8$
block being filtered. When processing the pixel at position $(i,j)$, the
filter is allowed to use pixels $x(i+m,j+m)$ lying outside of the
$8\times8$ block. If
the input pixel lies outside of the frame (visible area), then the pixel
is ignored ($f(d,S,D)=0$). To maximize parallelism, CDEF always operates
on the input (post-deblocking) pixels $x(i,j)$ so filtered pixels are never reused
for filtering other pixels.

\subsection{Valid strengths and damping values}

The strengths $S^{(p)}$ and $S^{(s)}$ and damping $D$ must be set high
enough to smooth out coding artifacts, but low enough to avoid
blurring details in the image.  For 8-bit content $S^{(p)}$ ranges between $0$ and $15$, and $S^{(s)}$ can be $0$, $1$,
$2$ or $4$.  $D$ ranges from $3$ to $6$ for luma, and the
damping value for chroma is always one less.  $D$ shall
never be lower than the $\log_2 S$ to ensure that
the exponent of $2^{D-\lfloor{\log_{2}S}\rfloor}$ in (\ref{eq:constraint-function}) never
becomes negative.  For instance, if for chroma $S^{(p)}=15$ and the luma damping is $3$,
the chroma damping shall also be $3$ (and not 2) because $\left\lfloor\log_2 S^{(p)}\right\rfloor=3$.

For bit depths greater than 8 bits, $S^{(p)}$ and $S^{(s)}$ are scaled according to
the extra bit depth, and $D$ is offset accordingly. For example, 12-bit
content can have $S^{(p)}$ values of $0$, $16$, $32$, $...$, $240$, and
$D$ ranges from $7$ to $10$. The strengths are scaled up after selecting the
primary filter taps, so the taps still alternate, even though the scaling makes
all values of $S^{(p)}$ even.  Picking the optimal damping value is less
critical than picking the
optimal strengths.  $S^{(p)}$ and $S^{(s)}$ are chosen independently for
luma and chroma.

The signaled luma primary strength $S^{(p)}$ is adjusted for each $8\times8$
block using the \textit{directional\_contrast} value ($v$) computed in algorithm~\ref{cap:direction-search}:
\begin{equation}
S^{(p)}_{adj}=\left\{ \begin{array}{ll}
\left\lfloor\frac{S^{(p)} \left(4 + \min\left(\left\lfloor \log_2\left\lfloor\frac{v}{2^{16}}\right\rfloor\right\rfloor, 12\right)\right) + 8}{16}\right\rfloor &, v \ge 2^{10}\\
0 &, \mathrm{otherwise}
\end{array}\right. \label{eq:strength-adjustment}
\end{equation}
The adjustment makes the filtering adapt to the amount of directional contrast
and requires no signaling.

\section{Signaling and Filter Blocks}
\label{sec:signaling}

The frame is divided into filter blocks of $64\times64$ pixels.
Some CDEF parameters are signaled at the frame level, and some
may be signaled at the filter block level.  The following is
signaled at the frame level: the damping $D$ (2 bit), the number of
bits used for filter block signaling (0-3, 2~bits), and a list of 1, 2,
4 or 8 \textit{presets}.  One preset contains the luma and chroma primary strengths (4
bits each), the luma and chroma secondary
strengths (2~bits each), as well as the luma and chroma \textit{skip
condition bits}, for a total of 14 bits per
preset.  For each
filter block, 0 to 3 bits are used to indicate which preset is used.
The filter
parameters are only coded for filter blocks that have some coded residual.
Such ``skipped'' filter blocks have CDEF disabled.
In filter blocks that do have some coded residual, any $8\times 8$ block with
no coded residual also has filtering disabled unless the skip condition bit is set
in that filter block's preset.

Since the skip condition flag would be redundant in the case when both
the primary and secondary filter strengths are $0$, this combination
has a special meaning.  In that case, the
block shall be filtered with a primary filter strength equal to $19$,
a secondary filter strength equal to $7$, and the skip condition still set.

When the chroma subsampling differs horizontally and vertically,
e.g., for 4:2:2 video, the filter is disabled for chroma, and the chroma
primary strength, the chroma skip condition flag and the chroma
secondary strength are not signaled.

\section{Encoder Search}
\label{sec:encoder-search}

\begin{table}
	\caption{CDEF Bj{\o}ntegaard-delta~\cite{Testing-draft} rate for
		the objective-1-fast test set in AWCY. The AV1 and Thor encoders were
		tested for a high-latency (HL) configuration, a real-time, low-latency (LL)
		configuration, as well as for low latency and low-complexity (LL+LC)\label{tab:bd-rate}.}
	\centering{%
		\setlength\tabcolsep{2.1pt}
		\begin{tabular}{cccccc}
			\hline 
			\small{Encoding} & \small{PSNR} & \small{CIEDE} & \small{PSNR-HVS} & \small{SSIM} & \small{MS-SSIM}\tabularnewline
			\hline 
			\small{AV1 HL} & -1.08\% & -2.11\% & -0.15\% & -1.11\% & -0.44\%\tabularnewline
			\small{AV1 LL} & -1.93\% & -2.88\% & -0.86\% & -1.96\% & -1.18\%\tabularnewline
			\small{AV1 LL + LC} & -3.68\% & -4.54\% & -2.50\% & -4.15\% & -3.05\%\tabularnewline
			\hline
			\small{Thor HL} & -2.26\% & -3.13\% & -0.49\% & -2.75\% & -1.39\%\tabularnewline
			\small{Thor LL} & -3.19\% & -5.18\% & -1.34\% & -3.31\% & -2.23\%\tabularnewline
			\small{Thor LL + LC} & -6.17\% & -10.33\% & -4.13\% & -7.60\% & -6.11\%\tabularnewline
			\hline 
	\end{tabular}}
\end{table}

\begin{figure}
\centering{\includegraphics[width=\columnwidth]{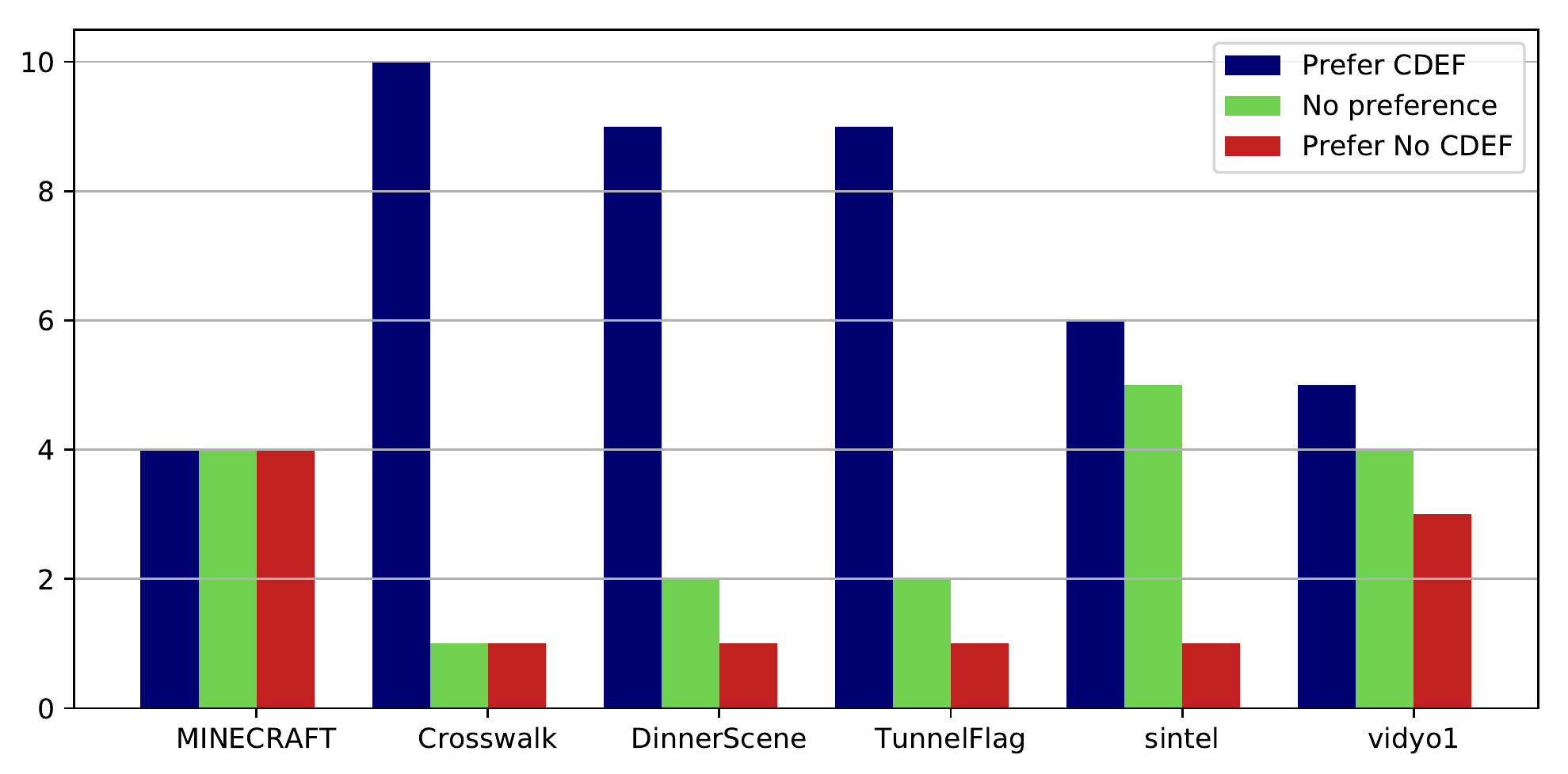}}
\vspace{-1.5em}
\caption{Subjective A-B comparison results (with ties) for CDEF vs. no processing for the high-latency configuration.\label{fig:subjective}}
\end{figure}

On the encoder side, the search needs to determine both the frame level parameters
(preset parameters, number of presets) and the filter block-level preset ID. Assuming
the presets are already chosen, the ID for each non-skipped filter block is
chosen by minimizing a distortion metric over the filter block. The simplest error metric
is the sum of squared error (SSE), defined as $D=\left\Vert\mathbf{s}-\mathbf{d}\right\Vert^2$,
where $\mathbf{s}$ is a vector containing the source (uncoded) pixels for the filter block and
$\mathbf{d}$ contains the decoded pixels, filtered using a particular preset. While SSE leads
to good results overall, it sometimes causes excessive smoothing in non-directional textured
areas (e.g. grass). Instead, we use a modified version of SSE that takes into account
contrast in a similar way to the structural similarity (SSIM) metric~\cite{wang2004image}. The
distortion metric is the sum over the filter block of the following $8\times8$ distortion function:
\begin{equation}
D_{8\times8} = \frac{\sigma_s^2 + \sigma_d^2 + C_1}{2\sqrt{\sigma_s^2\sigma_d^2 + C_2}}\cdot\left\Vert\mathbf{s}-\mathbf{d}\right\Vert^2\ ,\label{eq:cdef-dist}
\end{equation}
where $\sigma_s^2$ and $\sigma_d^2$ are the variances of $\mathbf{s}$ and $\mathbf{d}$ over the
block and the constants are set to $C_1=6.25$ and $C_2=312.5$ for 8-bit depth. Using (\ref{eq:cdef-dist}) degrades PSNR results, but improves visual quality.
Since the distortion metric is only used in the encoder, it is not normative.

There are many possible strategies for choosing the presets for the frame, depending
on the acceptable complexity requirements and whether the encoder is allowed to make
two passes through the frame. In the two-pass case, the first step is to measure the
distortion $D_{b,p}$ for each filter block b, for each combination $p$ of 
$S^{(p)}$, $S^{(s)}$ and skip condition bit
($16\times4\times2=128$ parameter combinations) and for each plane. 
Once the distortion values are computed, the goal is to find the two sets of presets
$P_{\mathrm{luma}}$ and $P_{\mathrm{chroma}}$ that minimize the rate-distortion cost
\begin{equation}
J=\lambda B\log_2 N +  \sum_{b} \left(\min_{p \in P_{\mathrm{luma}}} D^{\mathrm{luma}}_{b,p} + \min_{p \in P_{\mathrm{chroma}}} D^{\mathrm{chroma}}_{b,p}\right)\ ,
\end{equation}
where $N$ is the cardinality of $P_{\mathrm{luma}}$ and $P_{\mathrm{chroma}}$ and $B$ is
the number of filter blocks. While we are not aware of polynomial-time algorithms to find the
global minimum, we have found that a greedy search can produce near-optimal results. With
the greedy search, we start by finding the optimal presets for $N=1$ and then increment $N$
by finding the optimal preset to add while keeping the already-selected presets $1..N-1$ constant.
If the encoder can afford the complexity, it is possible to improve on the purely
greedy search by iteratively re-optimizing one preset at a time.

The high complexity version of the search described above typically results in less than
1\% of the encoding time. Still, for low complexity operation, it is possible to reduce
the search complexity by only considering a subset of the 128 possible presets. 
This results in only a small loss ($<0.1\%$ BD-rate) in quality.

The damping value may be determined from
the quantizer alone, with larger damping values used for larger quantizers.

\section{Results}
\label{sec:results}

We tested CDEF with the Are We Compressed
Yet~\cite{AWCY} online testing tool using the AV1 and Thor~\cite{Bjontegaard2016,Davies2016} codecs, both
still in development at the time of writing. BD-rate results for PSNR, PSNR-HVS~\cite{PSHRHVSM}, CIEDE2000~\cite{luo2001development}, SSIM~\cite{wang2004image}, and MS-SSIM~\cite{wang2003multiscale} are shown in
Table~\ref{tab:bd-rate} for the objective-1-fast test set.  These
results reflect the gains in the codebases for git SHA's \texttt{e200b28}~\cite{AV1-git}
(8$^\mathrm{th}$ August 2017) and \texttt{b5e5cc5}~\cite{Thor-git} (21$^\mathrm{st}$ October 2017) for AV1 and Thor
respectively.

Subjective tests conducted on AV1 for the high-latency configuration
show a statistically significant ($p<.05$) improvement for 3 out of 6 clips,
as shown in Fig.~\ref{fig:subjective}. Considering that it usually takes in the order
of 5\% improvement in BD-rate to obtain such statistically significant results, and the
tested configuration is the one with the smallest BD-rate improvement, we
believe the visual improvement is higher than the BD-rate results suggest. 

The objective results show that CDEF performs better when encoding with fewer tools and
simpler search algorithms. In some sense, CDEF ``competes''
for the same gains as some other coding tools. Considering that CDEF is significantly
less complex to encode than many of the AV1 tools, it provides a good way of reducing
the complexity of an encoder. 
In terms of decoder complexity, CDEF represents between 3\% and 10\% of the AV1 decoder
(depending on the configuration).

\section{Conclusion}

We have demonstrated CDEF, an effective in-loop filtering algorithm for removing coding
artifacts in the AV1 and Thor video codecs. The filter is able to effectively remove
artifacts without causing blurring through a combination of direction-adaptive filtering
and a non-linear filter with signaled parameters. Objective
results show a bit-rate reductions up to 4.5\% on AV1 and 10.3\% on Thor.
These results are confirmed by subjective testing.

CDEF should be applicable to other video codecs as well as image codecs. In the case of AV1,
an open questions that remains is how to optimally combine its search with that of ``Loop Restoration''~\cite{Mukherjee2017}, another in-loop enhancement filter in AV1.

\section{Acknowledgments}

We thank Thomas Daede for organizing the subjective test.

\bibliographystyle{IEEEtran}
\bibliography{daala}

\end{document}